\documentclass[aps,pra,twocolumn,groupedaddress,showpacs,preprintnumbers,amsmath,amssymb]{revtex4}
\usepackage{graphicx}
\usepackage{dcolumn}
\usepackage{bm}
\usepackage{amsmath} 
 
\begin{document} 

\title{Critical frequency for vortex nucleation in Bose-Fermi mixtures
in optical lattices}

\author{M. Guilleumas, M. Centelles, M. Barranco, R. Mayol, and M. Pi} 

\affiliation{Departament d'Estructura i Constituents de la Mat\`eria,
Facultat de F\'{\i}sica, \\
Universitat de Barcelona, 
Diagonal {\sl 647}, {\sl 08028} Barcelona, Spain}


\begin{abstract}
We investigate within mean-field theory the influence of a
one-dimensional optical lattice and of trapped degenerate fermions on
the critical rotational frequency for vortex line creation in a
Bose-Einstein condensate. We consider laser intensities of the lattice
such that quantum coherence across the condensate is ensured. We find
a sizeable decrease of the thermodynamic critical frequency for vortex
nucleation with increasing applied laser strength and suggest suitable
parameters for experimental observation. Since $^{87}$Rb--$^{40}$K
mixtures may undergo collapse, we analyze the related question of how
the optical lattice affects the mechanical stability of the system.
\end{abstract}


\maketitle


An essential feature of a rotating superfluid is the eventual
appearance of quantized vortices \cite{don91}. For this reason,
nucleation of vortices in dilute gases in rotating traps has been the
subject of intensive experimental and theoretical research (see e.g.\
Refs.\ \cite{dal96,fet01,theory1,theory2,theory3,theory4,experiment1,
experiment2,experiment3} and references quoted therein).

Motivated by recent advances in the manipulation of rotating
Bose-Einstein condensates (BEC) \cite{ros02,bre04} and of condensates
loaded into optical lattices \cite{had04}, and in anticipation of
future experiments, we study a slowly rotating BEC in a
one-dimensional (1D) optical lattice \cite{multivor}. Rotating
ultracold Bose gases in optical lattices are suitable systems to study
quantum phenomena. In particular, the excitations, stability and
dynamics of a vortex line in the superfluid regime have been recently
addressed \cite{vortex1OL}. Concurrently, the physics of degenerate
Bose-Fermi mixtures in optical lattices is attracting conspicuous
attention. Theoretical progress is being made in the basic
understanding of the system's quantum phase diagram and the superfluid
to Mott-insulator transition \cite{BFinOL}, and experimentalists have
already prepared a degenerate mixture of Rb and K atoms in a tight
optical lattice \cite{mod03}.

The aim of the present work is twofold. Firstly, to address the
formation of vortices in a coherent array of Bose-Einstein condensates
in a 1D optical lattice, and to determine the dependence of the
thermodynamic critical angular velocity $\Omega_c$ for vortex
formation \cite{dal96,fet01} on the laser intensity. Secondly, to
investigate the effect of trapped fermions on $\Omega_c$. Attractive
Bose-Fermi mixtures may experience collapse and, thus, we also discuss
the impact of the optical lattice on the mechanical stability of the
mixture.

We consider a zero temperature mixture made of a $^{87}$Rb Bose
condensate (B) and a degenerate $^{40}$K Fermi gas (F). They are
confined by the axially-symmetric external potentials of a harmonic
magnetic trap and of a stationary periodic optical lattice modulated
along the $z$-axis. The lattice is produced by a far detuned laser,
which hinders the possibility of spontaneuous scattering and yields
practically equal Rb and K optical potentials. The resulting potential
for each kind of atom $q=B,F$ is
\begin{equation}
V_q = \frac{1}{2}\,m_q (\omega_{q \perp}^2 r^2 + \omega_{q z}^2 z^2)+
        \frac{V_0}{2} \cos (2\pi z / d) \,,
\label{vextBF}
\end{equation}
where $m_q$ is the atomic mass. The radial and axial frequencies of
the harmonic trap are taken from a recent experiment \cite{had04}:
$\omega_{B \perp}= 2 \pi \nu_{B \perp}$ with $\nu_{B \perp}= 74$ Hz and
$\omega_{B z}= 2 \pi \nu_{B z}$ with $\nu_{B z}= 11$ Hz for $^{87}$Rb,
while those for $^{40}$K are a factor $(m_B/m_F)^{1/2} \simeq 1.47$
larger. The optical potential is determined by its period $d=2.7 \mu$m
\cite{had04} and depth $V_0=s E_R$, where $E_R=\hbar^2 \pi^2/2md^2=h
\times 80$ Hz is the recoil energy and $s$ is a dimensionless
parameter.

For $3 \times 10^5$ atoms of $^{87}$Rb confined in the harmonic trap,
the condensate is cigar-shaped with a Thomas-Fermi length $L_{\rm
TF}=84 \,\mu$m, and a radius $R_{\rm TF}= 6 \, \mu$m. The superimposed
optical potential splits the BEC over $L_{\rm TF}/d \sim 30$ wells. We
shall be concerned with a range of laser intensities that ensures
full coherence of the condensate across the whole system
\cite{vortex1OL}. The criterion for the Mott transition \cite{zwe03}
for the considered parameters leads to an estimate of a maximum
lattice depth for the superfluid regime of $V_0 \sim 150 \, E_R \sim h
\times 12$ kHz \cite{had04}, when quantum tunneling between
consecutive wells is still sufficient to retain coherence.

Within our mean-field approach, neglecting $p$-wave interactions,
the energy density functional that describes the boson-fermion mixture
at zero temperature with a singly quantized vortex in the condensate
along the $z$-axis has the form \cite{ro02,jez04}
\begin{eqnarray}
{\cal E}(\textbf{r}) &=& \frac{ \hbar^2 }{2 m_B} 
        (\mbox{\boldmath$\nabla$} n_B^{1/2})^2+
        \frac{ \hbar^2}{2 m_B}  \frac{1}{r^2} n_B +
        V_B \,n_B + \frac{1}{2} g_{BB}\, n_B^2
       \nonumber\\[1mm]
       & & \null + g_{BF} \,n_F\, n_B +
           \frac{\hbar^2}{2 m_F} \tau_F
       + V_F \,n_F \,,
       \label{ed}
\end{eqnarray}
where $n_B= |\Psi|^2$ is the condensate density, $n_F$ is the fermion
density, and $\tau_F$ is the fermion kinetic energy density written as
a function of $n_F$ and its gradients [Thomas-Fermi-Weizs\"acker (TFW)
approximation]. The boson-boson and boson-fermion coupling constants
$g_{BB}$ and $g_{BF}$ are written in terms of the $s$-wave scattering
lengths $a_B$ and $a_{BF}$ as $g_{BB}=4\pi\,a_{B}\hbar^2/m_B$ and
$g_{BF}=4\pi\,a_{BF}\hbar^2/m_{BF}$, respectively, with $m_{BF} \equiv
2 m_B m_F / (m_B + m_F)$.

The variation of ${\cal E}$ with respect to $n_B$ and $n_F$, under
the constraint of given number of bosons $N_B$ and fermions $N_F$,
yields two coupled Euler-Lagrange equations: a Gross-Pitaevskii
equation for bosons with a term describing the boson-fermion
interaction, and a TFW equation for fermions \cite{ro02,jez04}, which
is increasingly valid the larger $N_F$ is. Starting from randomly
sampled densities, we solve these equations as indicated in Ref.\
\cite{bar03}. We use the adapted version of the virial theorem
\cite{jez04} to check the numerical convergence.

\begin{figure}
\includegraphics[width=0.94\columnwidth,
angle=270, clip=true]{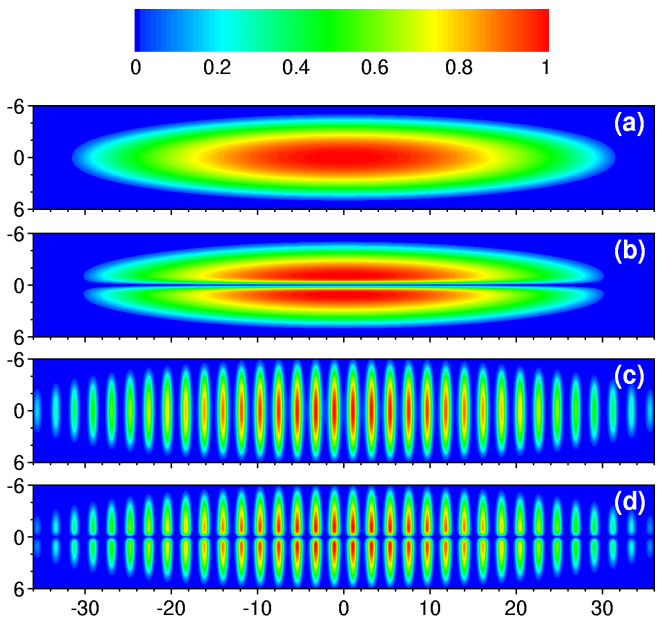}
\caption[]{(Color online)
Contour plots of the $^{87}$Rb condensate density in the $xz$-plane.
(a) Condensate in the pure harmonic trap.
(b) Same as (a) with a quantized vortex along the $z$-axis.
(c) Same as (a) under an optical lattice with $V_0/h =2.5 \,$kHz. 
(d) Same as (c) with a quantized vortex along the $z$-axis.
Distances are in units of the oscillator length 
$a_{B \perp}= \sqrt{\hbar/(m_B \omega_{B \perp})} = 1.25 \mu$m.
The density scale from 0 to 1 at the top of the figure is normalized
to the highest density in each panel [these are 229 in (a), 212 in
(b), 394 in (c), and $376 \,\, a_{B \perp}^{-3}$ in (d)].}
\label{fig1}
\end{figure}

We first consider a pure BEC, with $N_B=3 \times 10^5$ atoms of
$^{87}$Rb as in the experiment of Ref.~\cite{had04}. For the
boson-boson scattering length we use the value $a_B=98.98 \, a_0$ ($1
\, a_0 = 0.529 \,\mbox{\AA}$) \cite{kem02}. In Fig.~\ref{fig1} we show
contour plots of the condensate density in the $xz$-plane,
$n_B(x,0,z)=|\Psi(x,0,z)|^2$, for different situations. In (a) the
condensate is confined only by the harmonic trap. In (b) the
cigar-shaped condensate of (a) hosts a quantized vortex line along the
$z$-axis. In (c) a 1D optical lattice with $V_0/h = 2.5 \,$kHz splits
the cigar-shaped condensate (a) into an array of multiple disk-like
coherent condensates. As shown in Ref.\ \cite{kra02}, certain aspects
of the macroscopic properties and low-energy dynamics of a
magnetically trapped BEC in a tight optical lattice can be understood
in terms of a renormalized interaction coupling constant $g^*_{BB} >
g_{BB}$ and of an effective mass $m^*_B > m_B$ along the direction of
the periodic optical potential. The increase of the radial size of the
sample that one observes in panel (c) with respect to panel (a) is due
to the increased repulsive effect of the boson-boson interaction when
the system feels the optical lattice. It originates from the fact that
the optical confinement produces a local compression of the gas inside
each well \cite{zwe03,kra02}. The axial size of the condensate also
increases appreciably due to the combined effect of the repulsive
interactions and the redistribution of atoms inside the potential
wells. 

In panel (d) of Fig.~\ref{fig1} the optical lattice is superimposed to
the condensate with the vortex state (b). In a bulk superfluid, the
size of the vortex core is of the order of the healing length $\xi=[8
\pi n_B a]^{-1/2}$, where $n_B$ is the bulk density. This expression
holds for inhomogeneous superfluids taking for $n_B$ the local boson
density in the absence of vortices \cite{fet01}. Thus, the vortex core
size is larger for the outer sites of the split condensate, where the
density is smaller.

\begin{figure}
\includegraphics[width=0.95\columnwidth,
angle=0, clip=true]{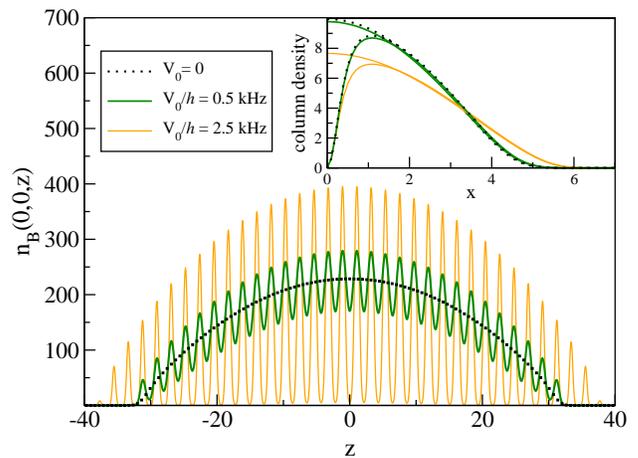}
\caption[]{(Color online)
Density profile of the condensate along the $z$-axis
$n_B(0,0,z)=|\Psi(0,0,z)|^2$, in units of $a_{B \perp}^{-3}$, for
different laser intensities. 
The coordinate $z$ is in units of $a_{B \perp}$.
Inset: column density $n_B(x)$, in units
of $10^3 \times a_{B \perp}^{-2}$, for the same $V_0$ values. The
column density of the condensate when it hosts a vortex line is also
plotted. The coordinate $x$ is in units of $a_{B \perp}$.
}
\label{fig2}
\end{figure}

Figure~\ref{fig2} shows the evolution of the density profile of the
condensate along the $z$-axis, $n_B(0,0,z)=|\Psi(0,0,z)|^2$, with the
laser intensity. In the presence of a shallow 1D optical lattice, the
condensate profile starts developing small oscillations around the
profile at $V_0=0$ which follow the periodic optical potential, but
its spatial extension is nearly not affected. When the laser intensity
increases the system nearly splits into separate condensates. The
effect of the local compression inside each potential well of the
lattice, as well as the increase of the axial size of the system, can
be clearly seen in Fig.~\ref{fig2}. In the inset we plot the column
density $n_B(x)=\int dz \, n_B(x,0,z)$ for the condensate in the
harmonic trap alone and when the optical lattice is added. We also
display $n_B(x)$ for the condensate hosting a vortex line, in this
case the quantized circulation around the vortex line pushes the atoms
away from it.

\begin{figure}
\includegraphics[width=0.95\columnwidth,
angle=0, clip=true]{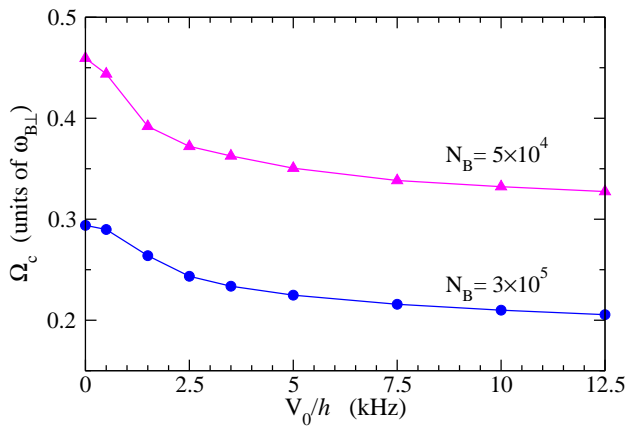}
\caption[]{(Color online)
Critical angular velocity $\Omega_c$ for nucleation of a singly
quantized vortex line as a function of the laser intensity $V_0$.
}
\label{fig3}
\end{figure}

The thermodynamic critical angular velocity $\Omega_c$ for nucleating
a singly quantized vortex is obtained by subtracting from the vortex
state energy $E_1$ in the rotating frame the ground-state energy
$E_0$, i.e., $\Omega_c=(E_1/N_B -E_0/N_B)/\hbar$ \cite{dal96,fet01},
and it provides a lower bound to the critical angular velocity
\cite{fet01}. In Fig.~\ref{fig3} we plot $\Omega_c$ as a function of
the laser intensity for $N_B= 3 \times 10^5$ and for $N_B= 5 \times
10^4$. For a fixed number of condensate atoms, $\Omega_c$ decreases
when $V_0$ increases, in agreement with the fact that the radial size
of the system becomes larger in the presence of the optical lattice
and the associated reduction of atoms along the symmetry axis (see the
decrease of the column density $n_B$ at the origin in Fig.~\ref{fig2}
with increasing $V_0$). In turn, the increase of the inertia of the
condensate along the direction of the laser beam due to the larger
effective mass $m^*_B$ \cite{kra02} also contributes to the reduction
of the critical angular velocity.

Assuming that the relative effect remains of the same order in
experiment, Fig.~\ref{fig3} predicts a sizeable $24\%$ reduction
of $\Omega_c$ already with a laser strength $V_0/h=5$ kHz for both of
the $N_B$ values. The effect is enhanced the shallower the magnetic
trap is. A less elongated magnetic trap would favor the appearence of
a vortex at a slower rotation. For example, the values of 
$\Omega_c(V_0=0)$ shown in Fig.~\ref{fig3} would be decreased by a 0.6
factor in a spherical trap with $\nu_{B z}= \nu_{B \perp}= 74$ Hz.
The relative reduction of $\Omega_c$ caused by $V_0/h=5$ kHz
with respect to $V_0=0$ would still be of 20--23\%.

Figure~\ref{fig3} suggests an experiment to test the dependence of the
critical frequency for vortex nucleation on the intensity of the
optical lattice, since it shows that, for a rotating condensate with a
given $N_B$, a vortex line should nucleate at angular frequencies
lower than $\Omega_c(V_0=0)$ if a co-rotating 1D optical lattice is
superimposed,  whereas otherwise it would not. As the thermodynamic
$\Omega_c$ underestimates the actual critical frequency for vortex
creation, dynamical calculations are needed for a precise quantitative
estimate of this effect.

\begin{figure}
\includegraphics[width=0.94\columnwidth,
angle=270, clip=true]{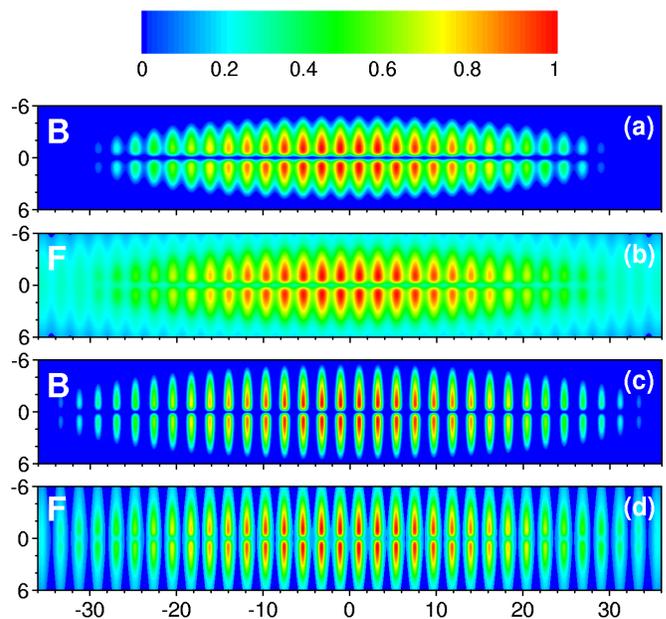}
\caption[]{(Color online)
Contour plots of the condensate (B) and fermionic (F) densities in the
$xz$-plane for the $^{87}$Rb--$^{40}$K mixture, in the combined
harmonic and optical lattice trap. The condensate hosts a quantized
vortex. In (a) and (b) $V_0/h = 0.5 \,$kHz, whereas in (c) and (d)
$V_0/h = 2.5 \,$kHz. Distances are in units of $a_{B \perp}$.
The density scale from 0 to 1 at the top of the figure is normalized
to the highest density in each panel [these are 312 in (a), 25 in (b),
457 in (c), and $37 \,\, a_{B \perp}^{-3}$ in (d)].}
\label{fig4}
\end{figure}

With the purpose of assessing the effects of a trapped fermion cloud
on the rotating condensate in an optical lattice, we consider next a
$^{87}$Rb--$^{40}$K mixture. It is characterized by a large attractive
Bose-Fermi interaction that assists the sympathetic cooling of the
fermionic species down to the degenerate regime. There are various
values of the interspecies scattering length $a_{BF}$ published,
depending on the type of experiment and techniques used
\cite{fer02,mod02,modcol03,gol04}. A recent measurement of $a_{BF}$ at
JILA \cite{gol04} has established a value $a_{BF}=-250 \pm 30 \, a_0$,
which we adopt here because of its reduced experimental uncertainties.
Vortex states in $^{87}$Rb--$^{40}$K mixtures confined by a harmonic
potential have been recently addressed \cite{jez04} (although
$a_{BF}=-395 \, a_0$ \cite{modcol03} was employed). When a 1D optical
lattice is switched on, the atomic species experience the trapping
potential (\ref{vextBF}).

We have carried out calculations for a mixture with $N_B=3 \times
10^5$ and $N_F=1.5 \times 10^5$ atoms, and the same parameters as in
Fig.~\ref{fig1}. We assume that the Fermi component is in the
normal---nonsuperfluid---but quantum degenerate phase, and consider
that it is in a stationary state. This situation could be achieved
experimentally by waiting long enough after the generation of the
vortex in the condensate to let the drag force between bosons and
fermions to dissipate.
We present in Fig.~\ref{fig4} contour plots of the
boson $n_B(x,0,z)$ and fermion $n_F(x,0,z)$ densities when the
condensate hosts a vortex line, for two laser strengths: $V_0/h = 0.5$
(a,b) and $2.5$ kHz (c,d). The condensate and the fermionic cloud are
modulated by the regular pattern of the optical lattice. The large
mutual attraction between fermions and bosons makes the effective
interaction between bosons less repulsive than in the pure BEC.
Furthermore, the boson atoms induce an effective attraction between
fermions counteracting the Fermi pressure. As a result, the density of
both species increases in the overlapping region, as if they were more
strongly confined by the external potentials, and the condensate
becomes more compact in space (compare Figs.~\ref{fig4}c and
\ref{fig1}d). Still, the effect of the Pauli exclusion principle is
notorious in Fig.~\ref{fig4} where the $^{40}$K cloud is seen to
extend to larger distances than the $^{87}$Rb atoms, both axially and
radially. 

The presence of a vortex along the $z$-axis in the condensate
component is apparent in panels (a) and (c) of Fig.~\ref{fig4} as the
boson density vanishes on the vortex line. It is interesting to note
that the large attractive $a_{BF}$ leads to a visible depletion of the
fermionic  density on the $z$-axis as well, reminiscent of the bosonic
vortex  core. The effect should be directly observable experimentally.
It is more evident in (d) where bosons and fermions are more tightly
confined by the optical lattice.

In the present mixture we find that the presence of the quantum
degenerate fermionic cloud does not change much the value of
$\Omega_c$ compared to the pure BEC, nor its pattern against the
optical lattice strength. Indeed, for all the $V_0$ values considered
here the critical frequency for vortex appearence is raised by
$\sim 10\%$, due to the enhancement of the condensate density at the
core caused by the fermion atoms. The impact of the fermion cloud on
$\Omega_c$ is magnified in a deeper magnetic trap. For instance, if
the same mixture was set in a trap with $\nu_{B z}= 95$ Hz and $\nu_{B
\perp}= 640$ Hz, that preserves the aspect ratio $\nu_{B z}/\nu_{B
\perp}=0.15$ used in Fig.~\ref{fig4}, the critical frequency
$\Omega_c(V_0=0)$ would increase by $40\%$ with respect to the $N_F=0$
case. In this squeezing magnetic trap, however, the effect of $V_0$ on
$\Omega_c$ is negligible ($<2\%$).

A peculiar feature of the $^{87}$Rb--$^{40}$K system, stemming from
the interplay of the moderately repulsive $a_B$ and the strongly
attractive $a_{BF}$, is the existence of a mechanical stability limit
beyond which the system cannot sustain more fermions, for at high
densities the strong interspecies attraction may overcome the Pauli
pressure and drive the ultracold gas mixture into collapse. The
phenomenon has been observed experimentally \cite{mod02}. Thus, we 
next address the stability of the $^{87}$Rb--$^{40}$K system
subject to an optical lattice.
In fact, collapse is
currently the focus of experimental attention because of the
implications the instability has for constraining $a_{BF}$
\cite{mod02,modcol03,gol04}, and also because there are
prospects that a radially-squeezed $^{87}$Rb--$^{40}$K mixture at a
density close to collapse must be able to form stable bright soliton
trains \cite{bon04}.

Keeping $N_B$ fixed at $3 \times 10^5$ atoms as in the previous
discussions, with $a_{BF}=-250 \, a_0$ we find that the boson-fermion
mixture confined in the magnetic trap of Fig.~\ref{fig4} would be
stable up to virtually arbitrary fermion numbers, consistently with
the findings of Ref.~\cite{ro02}. Increasing $\omega_{B \perp}$
prompts the occurrence of collapse. In a trap with $\nu_{B z}= 95$ Hz
and $\nu_{B \perp}= 640$ Hz, the $3 \times 10^5$ rubidium atoms are
able to retain up to a maximum of $N_F^{\rm max}= 2.6\times 10^5$
fermions, when the system collapses. (The effect on $\Omega_c$ of
$N_F=1.5\times 10^5$ and $V_0$ for this trap has been discussed
above.) Now, a decrease of $N_F^{\rm max}$ can be seen when a 1D
optical lattice is added, caused by the effective local compression of
the atoms inside the optical trap. Indeed, superimposing a 1D optical
lattice with $d=2.7 \mu$m and $V_0/h= 5$ kHz to this harmonic trap,
the $^{87}$Rb--$^{40}$K mixture with $N_B=3\times 10^5$ can sustain up
to $N_F^{\rm max}= 7.5\times 10^4$ fermions only, i.e., less than 30\%
of the $V_0=0$ value. The effect appears to be accessible to
verification under present experimental conditions. It opens the
possibility to study the collapse of a mixture for trapped fermion and
boson numbers considerably smaller than in a pure harmonic trap.

In conclusion, we have determined the thermodynamic critical frequency
for nucleating a quantized vortex in an array of coherent condensates
within the Gross-Pitaevskii theory. We have studied a Bose-Fermi
mixture trapped in a 1D optical lattice to examine the effect on the
value of $\Omega_c$ of a fermionic cloud in the quantum degenerate
phase. $^{87}$Rb--$^{40}$K mixtures may collapse and, in this regard,
it has been shown that a 1D optical lattice can be an efficient degree
of freedom to tune the onset of instability. Our analysis of the
relative variation of $\Omega_c$ calculated by thermodynamic arguments
is intended to provide a useful information for future experiments. A
quantitative determination of $\Omega_c$ requires dynamical
calculations. Work in this line is being undertaken.

\acknowledgments{We thank Dr.\ K. Bongs for valuable discussions. This
work has been performed under Grants No.\ BFM2002-01868 from DGI
(Spain) and FEDER, and No.\ 2001SGR-00064 from Generalitat de
Catalunya. M.G. thanks the ``Ram\'on y Cajal'' Program (Spain) for
financial support.}

\end{document}